# The hemagglutinin mutation E391K of pandemic 2009 influenza revisited

Jan P. Radomski [1,4,*], Piotr Płoński [2], Włodzimierz Zagórski-Ostoja [3]

[1] *Interdisciplinary Center for Mathematical and Computational Modeling, Warsaw University, Pawińskiego 5A, Bldg. D, PL–02106 Warsaw, Poland*
[2] *Institute of Radioelectronics, Warsaw University of Technology, Nowowiejska 15/19, PL-00665 Warsaw, Poland*
[3] *Institute of Biochemistry and Biophysics, Polish Academy of Sciences, Pawińskiego 5A, Bldg. D, PL–02106 Warsaw, Poland*
[4] *Institute of Biotechnology and Antibiotics, Starościńska 5, PL–02516 Warsaw, Poland*

**PREPRINT**

## Abstract

Phylogenetic analyses based on small to moderately sized sets of sequential data lead to overestimating mutation rates in influenza hemagglutinin (HA) by at least an order of magnitude. Two major underlying reasons are: the incomplete lineage sorting, and a possible absence in the analyzed sequences set some of key missing ancestors. Additionally, during neighbor joining tree reconstruction each mutation is considered equally important, regardless of its nature. Here we have implemented a heuristic method optimizing site dependent factors weighting differently $1^{st}$, $2^{nd}$, and $3^{rd}$ codon position mutations, allowing to extricate incorrectly attributed sub-clades. The least squares regression analysis of distribution of frequencies for all mutations observed on a partially disentangled tree for a large set of unique 3243 HA sequences, along all nucleotide positions, was performed for all mutations as well as for non-equivalent amino acid mutations – in both cases demonstrating almost flat gradients, with a very slight downward slope towards the 3'-end positions. The mean mutation rates per sequence per year were $3.83*10^{-4}$ for the all mutations, and $9.64*10^{-5}$ for the non-equivalent ones.

## Keywords

phylogenetic analysis; influenza virus; hemagglutinin; neighbor joining; mutation rate; mutation E391K.

## 1. Introduction

Influenza viruses are antigenicaly variable pathogens, capable of continuously evading immune response. Accumulation of mutations in the antigenic sites is called the ''antigenic drift''. In circulating influenza viruses this antigenic drift is a major process, accumulating mutations at the antibody binding sites of receptor proteins, and enabling the virus to evade recognition by hosts' antibodies, which often translates into periodic epidemics of influenza. To tame the influenza spread a flexible vaccination WHO's program, based on periodic production of novel versions of vaccine, is adapted to the actually prevalent stain(s). For such programs the data on phylogenesis of circulating versions of pathogens, and genetic stability of their hemagglutinin (HA) sets data, could help to rationalize possible epidemiological measures. The rich collection of the data gathered recently during the last bout of the pandemic H1N1 flu offers possibility to give a fresh look on those perennial questions of influenza epidemiology. In March 2009 a novel influenza A virus H1N1 was identified in humans, initiating the first influenza pandemic of the 21st century. At the beginning of November 2009, there have been over 440,000 laboratory confirmed cases of pandemic influenza virus, resulting in more than 5,700 deaths worldwide. Nelson (Nelson et al. 2009) studied the early spread of H1N1 pandemic strains. At least 7 phylogenetically distinct viral clades have disseminated globally, and co-circulated in localities that experienced multiple introductions of pandemic H1N1. The role of founder effects is important for epidemiological scenarios (Lee et al. 2010) assuming that a genetic variability common to a small founder population will then also be found in most descendants. In viral outbreaks, such effects can be at play when specific mutations are enriched in samples coming from the same region, and/or same time. Considering phylogenetic relations it is useful to identify such viral lineage founder events. In the case of the 2009 pandemic influenza A, some mutations have received particular attention due to their apparent increased occurrence in many cases. One example of such new prominent mutation: the hemagglutinin HA-E391K (sometimes designated also as E374K, or E47K when counted at the position 47 in the HA2 region of hemagglutinin) was first described by Maurer-Stroh (Maurer-Stroh et al. 2010). They observed a rapid spread of strains carrying this mutation in the II-nd half of the 2009, and described its occurrence in the global, as well as in the local, Singaporean, surveillance data. According to their report the mutation HA-E391K, first identified in New York in July 2009, appeared shortly afterwards in Singaporean samples, and then grew rapidly to 90% locally by Dec. 2009 – in contrast to the presence in only 35% of global strains.

---

[*] corresponding author: janr@icm.edu.pl; voice +48-228749147.

  The spread of the E391K mutation was examined subsequently also in Finland (Ikonen et al. 2010, Strengell et al. 2010), Australia and New Zealand (Barr et al. 2010), Brazil (Ferreira et al. 2010), UK (Galiano et al. 2010), Canada (Graham et al. 2009), Italy (Piralla et al. 2012), Malaysia (Barlaj et al. 2011), and Japan (Obuchi et al. 2011). Mak *et al.* (Mak et al. 2011) undertook a detailed examination of temporal changes in the HA sequences carrying the mutation E391K from its earliest occurrence in early May 2009 till Jan. 2011 in Hong Kong. In Taiwan (Kao et al. 2012) strain carrying the E391K was first detected three weeks before the epidemic peak, evolved through the epidemic, and emerged finally as the major circulating strain, with significantly higher frequency in the post-peak period than in the pre-peak (64.65% *vs* 9.28%). The mutation persisted until ten months post-nationwide vaccination. Foremost spreading of the HA-E391K mutation has been associated with the fitness of the virus (Nelson et al. 2009, Mak et al. 2011, Tejeda et al. 2011). Hu (Hu 2011) has used the clustering affinity propagation to find typical exemplars of the HA sequences of the pandemic H1N1, and to probe for potential cooperative interactions between HA and neuraminidase. The exemplars, representing six clusters, formed the basis for determining nine key mutations in HA (including E391K). Such persistence suggest a need to weight out mutation frequencies along the whole HA gene, however, the task depends heavily on adequate phylogenic analysis, which must be exhaustively data-based. Due to its atypical features and epidemiological consideration, the H1N1 pandemic was followed by extensive screening effort, followed by sequencing of numerous isolates. This created the currently most extensive H1N1 sequence database with over 9000 accessions. Half of these complete gene sequences have precise isolation data labels, allowing detailed analyzes of the phylogeny of circulating variants, and facilitating frequency rates evaluation of HA mutations for this fast evolving and fast spreading virus.

The global strain sequencing efforts, combined with robust statistics allows novel insights into the phylogeny, and especially variability of this highly changeable RNA virus. The second problem is of high interest in view of recent results concerning the switching of receptor selection by the hemagglutinin (Imai et al. 2012, Herfst et al. 2012). Russell (Russel et al. 2012)], based on these combined experimental results, proposed a mathematical model of within-host H5N1 virus evolution to study some aspects influencing increase or decrease in probability of subsequent substitutions leading to aforementioned switch. They stressed that more data are needed for assessing calculated evolution rates based on the assumed mutation rates, which are of high interest for weighting out the speed of evolution of HA in switching receptor selection. In the current work we postulate that rates of mutation frequencies in HA commonly accepted are routinely overestimated by at least one order, and propose an enhanced method of finding evolutionary correlations between multiple strains of the H1N1 2009 pandemic virus.

## 2. Materials and Methods

As of June $9^{th}$ 2012 there were 9216 HA nucleotide sequences accessible in the NCBI influenza resource, of which 6118 were complete genes – of 1701 bases each. From this collection we have selected unique sequences only – such that from each subset of identical genes, the ones with the earliest dates of sample isolation were chosen as representatives, and a preference given to the records with the full *'month-day-year'* labels (in preference to records with only a *'month-year'*, or only a *'year'*, data) – forming the set of 3243 sequences analyzed here (0.96% with only the *'year'* time stamp, and 6.54% with only the *'month-year'* one).

Phylogenetic analyses were performed using the classic Neighbor Joining (Tamura et al. 2004, Waterhouse et al. 2009), our modified NJ algorithms: QPF (Płoński and Radomski 2010), and NJ+ (Płoński and Radomski 2013), as well as the approximately maximum-likelihood Fast-Tree method (Price *et al.* 2009). Also the results from the II-iteration trees after multiple alignment, available through the MUSCLE package (Edgar 2004), were checked for consistency with distance-based methods. For constructing maximum parsimony trees the DNApars program was used on the Mobyle server (Mobyle 2012), with its results augmented by our own script to identify, among multitude of equally probable maximum parsimony results, a tree requiring a smallest number of overall mutations to yield a minimal evolution graph of sequences in the analyzed set. For tree manipulations (in the Newick format) and their visualization the Dendroscope package (Huson et al. 2007) was used.

## 3. Results

Inference of evolutionary pathways from traditionally generated phylogenetic trees is a difficult task for two reasons: first, in traditional phylogenetic tree all sequences are shown as leaves – the internal nodes are not attributed to any particular sequences. We can interpret such a tree as showing only sequences that had no descendants. However, it is more complicated, because even if there are sequences, which have their descendants in the set, they still will be shown only as leaves. Thus, the construction of traditionally generated tree does not augment finding of evolutionary pathways and interpretation. Secondly, many phylogenetic methods generate only binary trees, with all the nodes having a degree three or less. It is significant constraint, as generally any ancestor can have any number of immediate descendants. For a sequence that has more than two descendants, all of them will be shown on such a tree as an assembly of auxiliary internal nodes and leaves. Binary character of traditional tree is a consequence of clustering algorithms, which always merge pairs of sequences. Keeping all these drawbacks in mind means that finding evolutionary pathways will grow to a problem of finding all ancestor-descendant pairs in a tree, which would require always running a check on all the nodes for every sequence – if anything, such an approach is computationally quite demanding. For these reasons we have recently proposed the QPF method (Płoński and Radomski 2010) to facilitate an unambiguous evolutionary pathway finding. Another problem, often present during phylogenetic analysis, is a possible absence of key sequences in the analyzed set.

All studies of the E391K mutation referred to in the Introduction were based on relatively small collections, of up to several hundred strains. Our preliminary analyses carried on the full set of the 3243 HA nucleotide sequences indicated, however, that while the E391K mutation is currently indeed a foremost one, the results depend rather significantly on the size of the analyzed set.

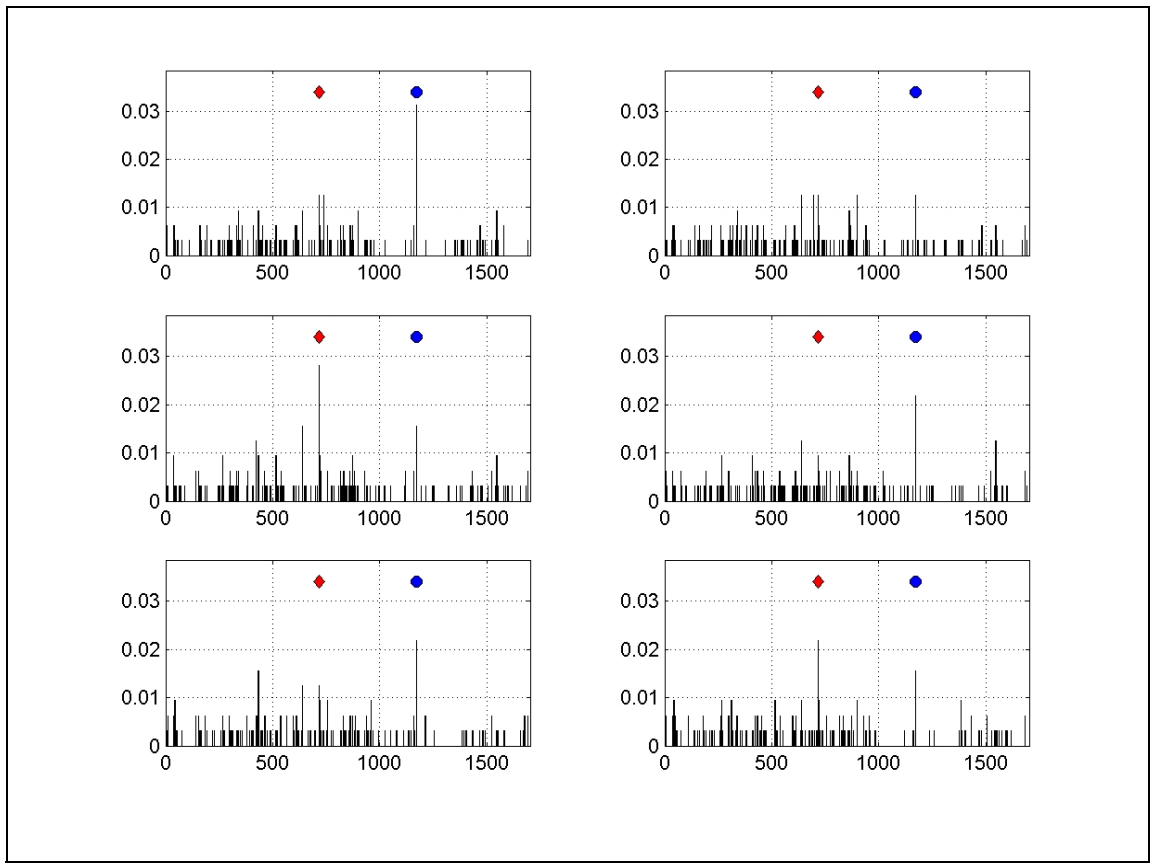

**Figure 1** – The comparison of non-equivalent amino acid mutation frequencies per nucleotide position of 2009 pandemic H1N1 hemagglutinin, obtained for several, randomly drawn, subsets of 320 sequences each (*see main text for details*), calculated on their Neighbor Joining trees. The red diamond marks the mutation A716G, whereas blue circle the G1171A one. The vertical axis corresponds to observed mutation frequencies normalized per sequence. The HA-1 epitopic region spans 987 nucleotides, starting from the position 49, so the mutation G1171A lies slightly outside of it.

The problem of reconstruction accuracy and its dependence on sampling rates is well recognized in phylogenetic analysis (Rannala et al. 1998), Two major underlying reasons are: the incomplete lineage sorting (ILS), and especially a possible absence in the analyzed sequences set some of key missing ancestors (MA). The problem of ILS received a lot of attention in the field of inferring species trees from gene trees (Leache et al. 2011, Edwards and Rausher 2009, Than and Nakhleh 2009). However, some issues leading to branches entanglement are also common in unraveling topology for a strain tree. In particular, when distances of sequence pairs are estimated within a set, it often happens that a given sequence might be less distant to some other sequence from an entirely different evolutionary pathway, than to its actual immediate ancestor, leading to its wrong branch assignment. The influence of MA would be especially evident while analyzing synthetic sequences sets – as the complete inheritance information contained therein can be unambiguously compared with resulting tree reconstructions (Płoński and Radomski 2010, Płoński and Radomski 2013, Ho and Jermiin 2004).

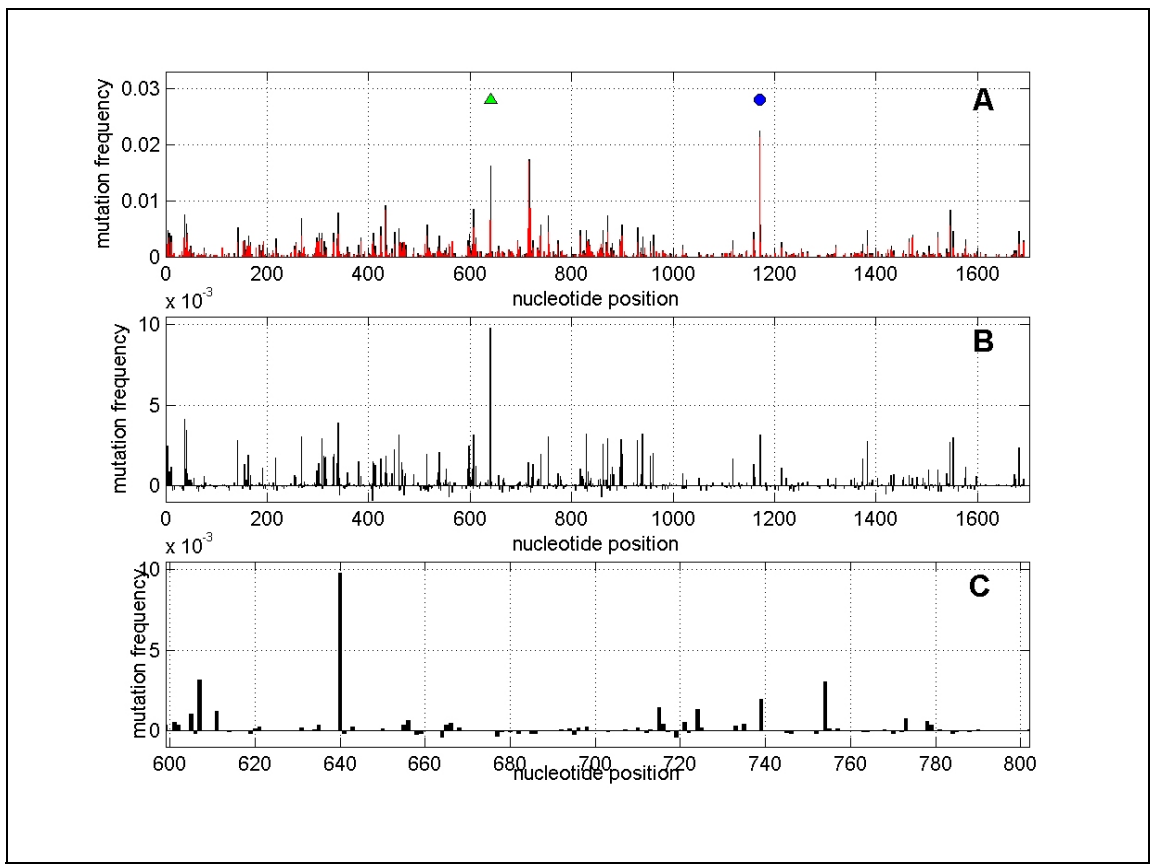

**Figure 2** – The comparison of non-equivalent aminoacid mutation frequencies per nucleotide position of 2009 pandemic H1N1 hemagglutinin: (panel **A**) – the respective results for the whole HA set of 3243 sequences (red bars), and the average frequencies from the 25 MC experiments (black bars; the green triangle marks the mutation G640A, whereas blue circle that of the G1171A one). The respective differences between the average mutation frequencies from 25-runs, and the corresponding values for the whole 3243 set (panel **B**). The enlarged inset from **B** for a clear view (panel **C**). The vertical axis corresponds to observed mutation frequencies normalized per sequence.

Two artifacts, both stemming from an increasing proportion of MAs in a set, are most troublesome. Artifact [a] concerns the orphaned sequences. When large fragments of evolutionary pathways are missing – leading to a decay of historical signal (Ho and Jermiin 2004) – then the only way to assign a pedigree of such an orphan is by linking it directly to a root of a whole tree (or sub-root of a branch).

Which in turn can lead to [b] an overestimation of mutation rates at some branching node positions. A possible ratio of MAs for the whole 3243 sequences HA set can be only roughly guess estimated, although the collection of 2009 pandemic isolates is perhaps one of the best in the recent history. However, by comparing the results for the whole set, with the corresponding data for its subsets, an influence of MA ratio on mutation rates can be illustrated rather well. To this end we have performed series of multiple Monte Carlo (MC) experiments on the 3423 set, by randomly selecting just 10% of sequences each time, and then analyzing each subset for the number of detected mutations. The **Figure 1** shows examples of mutation frequencies for the non-equivalent amino acids (as observed per nucleotide position, based on the change of character between four groups: non-polar, polar, acidic, and basic AA), carried on randomly drawn subsets (of the 320 different orthologs each time).

The E391K mutation corresponds to the G1171A (position marked with a blue circle on **Fig. 1**), with 2352 sequences in the set carrying 1171G before mutation, and 891 with 1171A afterwards. As the 891 subset forms the largest group of mutated sequences for any single position, the prominence of this mutation is not surprising. However, the variance between the mutation frequency distributions for the different random groups is quite striking – in some MC cases the E391K clearly predominates, in other sets another mutation, the A716G is more prominent, or neither of the two. The comparison of the average frequencies from the 25 MC experiments (panel **A**, black bars), and the respective results for the whole 3243 HA set (panel **A**, red bars) is shown on **Figure 2.** The frequency of the G1171A mutation was overestimated in the average of the 25 runs by about 5%, which might be considered a moderate error. However, the second most prominent mutation G640A was overestimated quite grossly, as the value for the whole set comprised only 39.9% of the mean from 25 runs. The panel **B** gives the respective differences between the averages of 25 runs, and the values for the whole 3243 set (the panel **C** is an inset from the **B** for a more clear view). Interestingly the A716G mutation, which was clearly dominant in some of the 25 MC trees (*cf.* the left-middle panel of **Fig. 1**, marked with red diamond), shows very minute difference in comparison with the whole set's frequency values – **Fig. 2C**. The choice of checking out the mutation rates on small 10% subsets was deliberate, as many of such estimates carried out in the past (Fitch et al. 1997, Bush et al. 1999, Ferguson et al. 2003) – and hence participating now to a high degree in the paradigm forming – were conducted on sets of comparable sizes. Fitch (Fitch et al. 1997) studied the 254 human influenza A/H3N2 virus genes, and found that the most parsimonious tree required 1,260 substitutions of which 712 were silent, and 548 were replacement substitutions – the HA1 domain of the HA gene was evolving at a rate of 0.0057 substitutions per site per year. Our estimates (*cf.* **Figs. 1** and **2**), however, show that the magnitude order of per site substitution rates is much lower. They reach a qualitative agreement with the 0.0018 averages estimated (Ferguson et al. 2003) for the H1N1 seasonal influenza, **only at peak values for the few sites** displaying the highest levels of mutability.

The overestimation of mutations frequencies analyzed for the 25 MC sequence sets, which all have very high ratio of MAs, is even more striking when also silent mutations are considered – this is shown on **Figure 3**: for the whole 3243 set (panel **A**), and for average from the 25 random experiments (panel **B**). It was hypothesized (Fitch et al. 1997, Bush et al. 1999) that fixation rates averaged over the whole gene length obscures variation among codons. The excess of non-synonymous versus synonymous substitutions at these very variable codons might reflect intense positive selection, possibly because the mutations are associated with the antigenic amino acids substitutions on the HA surface. However, our findings illustrate that there are relatively few regions along the whole 1701 nucleotide sequence for which there was not a single mutation observed.

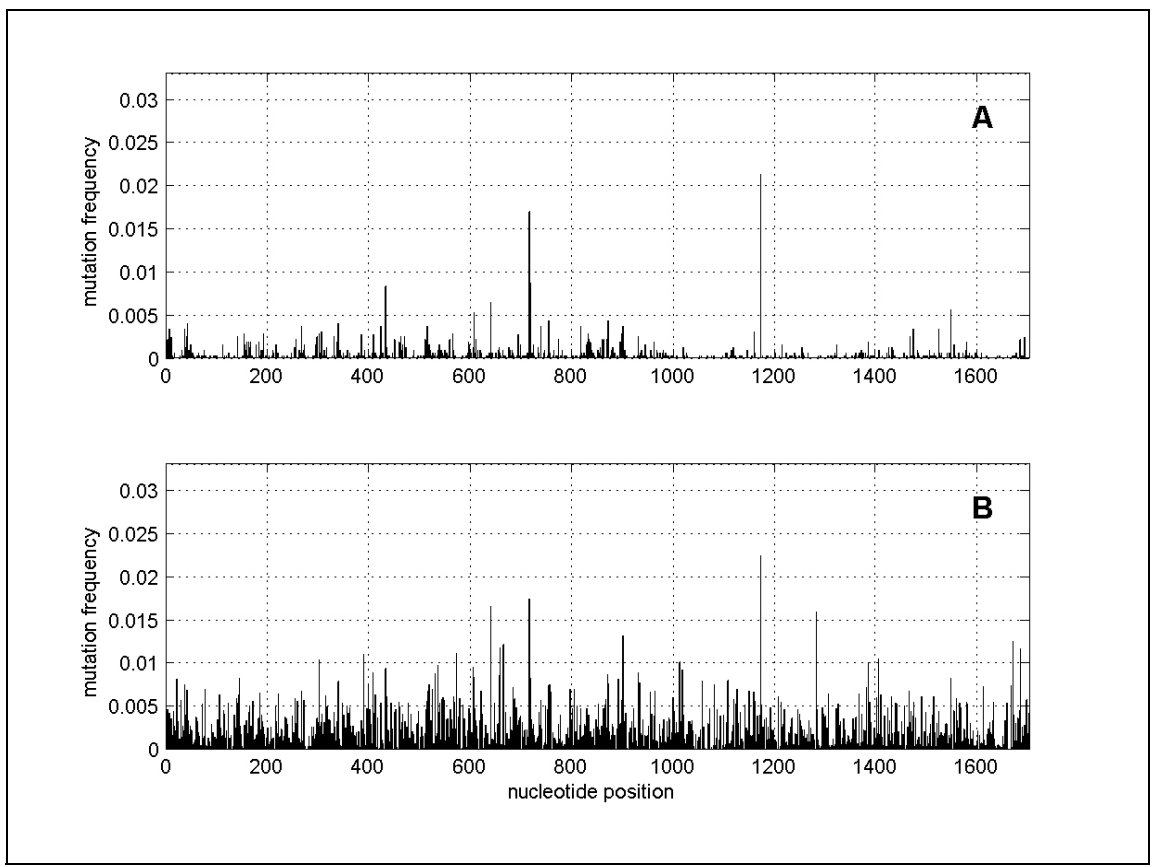

**Figure 3** – The comparison of all mutation frequencies per nucleotide position of 2009 pandemic H1N1 hemagglutinin. The respective results for the whole HA set of 3243 sequences (panel **A**), and the corresponding average frequencies from the 25 MC experiments (panel **B**). The vertical axis corresponds to observed mutation frequencies normalized per sequence.

However, the presented analysis of overestimated mutation frequencies does not paint a whole picture. In the tree for the whole 3243 set there were, e.g., multiple occurrences of the E391K mutation**:** [a] the 113 cases of the G1171A substitution, and [b] 24 cases of the back A1171G one. Thus a rather obvious question arises – are the multiple mutations of the same type, observed at the same nucleotide positions in closely related sequences, a result of algorithmic artifacts stemming from some possible ILS and MA difficulties? Or maybe, for some yet unknown reasons, the evolution of the virus manifests itself – repeatedly exploring the same newly "discovered" recurring mutation over and over again at different moments in parallel descendancy paths? Therefore, we decided to check if, and to what extent, a founder effect might be responsible, and if so, whether it will be possible to find a better way of evolutionary pathways reconstruction. The whole 3243 tree can be divided into three parts**:** all the clades containing only G1171; all the clades containing only A1171; and finally all the remaining sub-clades containing repetitive G1171A mutation chains – involving 203 unique sequences altogether. Acting on this small subset of 203 strains, the maximum parsimony method produced 10000, equally probable, candidate solutions, which were then checked for trees requiring minimum number of G1171A mutations, and the final results were augmented by the QPF method (Płoński and Radomski 2010) to find the possible MAs. The best resulting tree thus obtained required just one SNP – between the strains**:** A/New_York/3210/2009 (CY041549, collected on April 24, 2009), and A/Scotland/EastKilbride_420874/2009 (CY107784, collected on June 24, 2009). The computational cost of known algorithms that guarantee solution of finding an optimal maximum parsimony tree increases exponentially with problem size; so the practical computational considerations restrict their use to analyzing sequence sets of only relatively small size. Day (Day et al. 1986) established that the problem is NP-hard, and therefore so difficult computationally, that efficient optimal algorithms are unlikely to exist. But, if a single problem in NP-complete can be solved, then every other such problem in NP can also be reduced in a polynominal time, because by definition a problem in NP must be quickly reducible to every problem in NP-complete. Thus, by finding that such solution does exist, we received a momentum to search for a heuristic way – even if only approximately optimal – of finding trees that would possibly minimize the necessary number of explanatory mutations also across other than G1171A nucleotide positions.

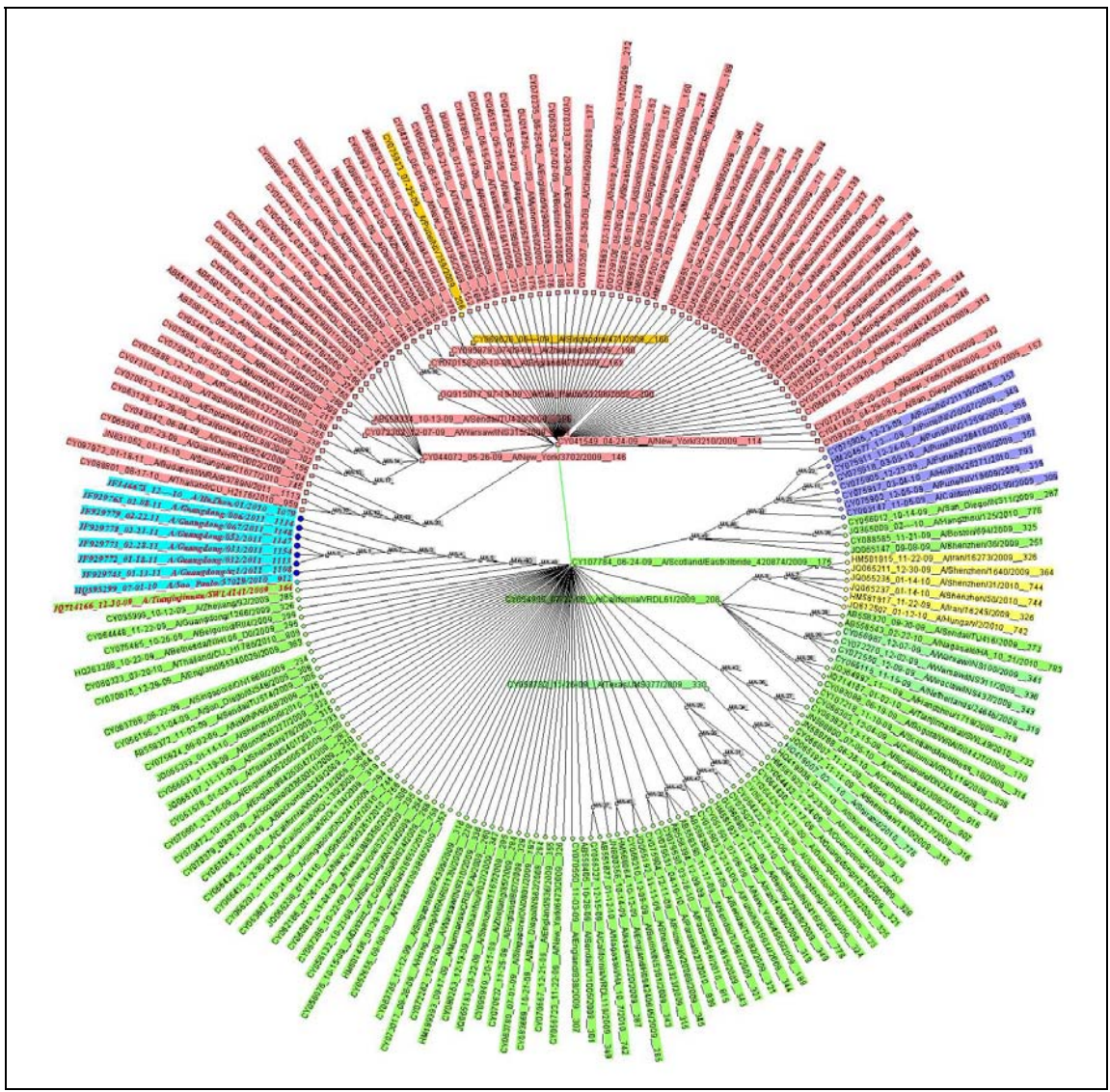

**Figure 4** – The NJ+ circular phylogram for the 2009 pandemic virus H1N1 hemagglutinin 203 sequences set (*see the main text for explanation of this set and the tree origin*). Each of sequence labels comprises of four parts, e.g. for the CY041549_04-24-09__A/New_York/3210/2009__114 there is the accession, isolation time stamp in the MM-DD-YY format, the strain name, and the number of days that passed since the Jan 1$^{st}$ 2009 till the day of sample isolation. Some labels of internal nodes are not displayed to preserve clarity. A zoomable Fig_4.pdf file in Supplementary materials should be used for a better visibility of details.

The NJ method is currently the most widely used technique for building phylogenetic trees using distance matrices. However, in the light of seemingly spurious artifacts of numerous G1171A mutations observed here, it should be noted that a very relevant question of how exactly NJ might influence the reconstruction of a gene's evolution pathways, have only recently been a subject of rigorous mathematical investigations (Gascuel and Steel 2006 for a review). A partial answer lies in the NJ algorithm's 'greedy' nature – as when it performs the search in a tree space, each step is guided predominantly by minimizing the global tree length criterion. However, NJ does not explore the whole tree space, and thus there is no guarantee of finding an optimally meaningful biological tree. In particular each mutation is considered equally important, independently of its nature (i.e. the silent ones, the amino acid changing or equivalent, the transversions or transitions – are all weighted the same). In an attempt to redress such a 'single mindedness' we have implemented a heuristic method (the details are given in the Appendix – *supplementary materials*), which optimize site dependent factors weighting differently 1$^{st}$, 2$^{nd}$, and 3$^{rd}$ codon position mutations. The results obtained are rather encouraging. A solution for the 203 sequences set is demonstrated on **Figure 4**. It can be seen that the E391K is indeed a predominant mutation between CY041549 and CY107784 (marked by a thick, green line)**:** all other G1171 sequences (marked red; *c.f.* also a non-disentangled NJ+ tree reconstruction for the same 203 set on **Figure S1** in *supplementary materials*) were found to be descendants of the New York strain, while all the other 1171A sequences (marked green) were descendants of the Scottish one. To illustrate, some smaller sub-clades (also disentangled) are marked as well**:** like the significant mutation on the first codon position C862T – changing Pro into Ser (P288S, marked cyan), or a co-occurring even smaller sub-clade of another amino acid changing first codon position, mutation A934G (I312V, emphasized by bigger, dark blue node markers). It is noteworthy that the whole 'cyan' sub-clade is relatively late (Dec. 2009, and the beginning of 2011), and therefore distant by many MAs from the CY107784, accumulating multiple silent mutations during that period. They included, among others**:** the T647C (Cys$_{215}$, shared with the A/Tianjinjinnan/SWL4141/2009 strain; marked by different typeface); G390A (Arg$_{130}$, accessions JF346675, JF929765, JF929772, JF929773, JF929778, JF929779); A522C (Ser$_{174}$, accessions JF929765, JF929772, JF929773, JF929778, JF929779); A801G (Val$_{267}$; accessions JF346675, JF929743, JF929765, JF929772, JF929773, JF929778, JF929779). Another example of a silent mutation is the C483T (Phe$_{161}$, the sub-clade marked violet). However, not always such full nodes disentanglement is possible, as different string positions might lead to contradictory sub-trees, which then can not be completely reconciled. An example of such situation is the A719G substitution – a silent mutation on the third position of the Ser$_{239}$ codon – forming two different sub-clades: one within the large 'green' cluster (marked yellow), and another (marked orange) within the 'red' cluster.

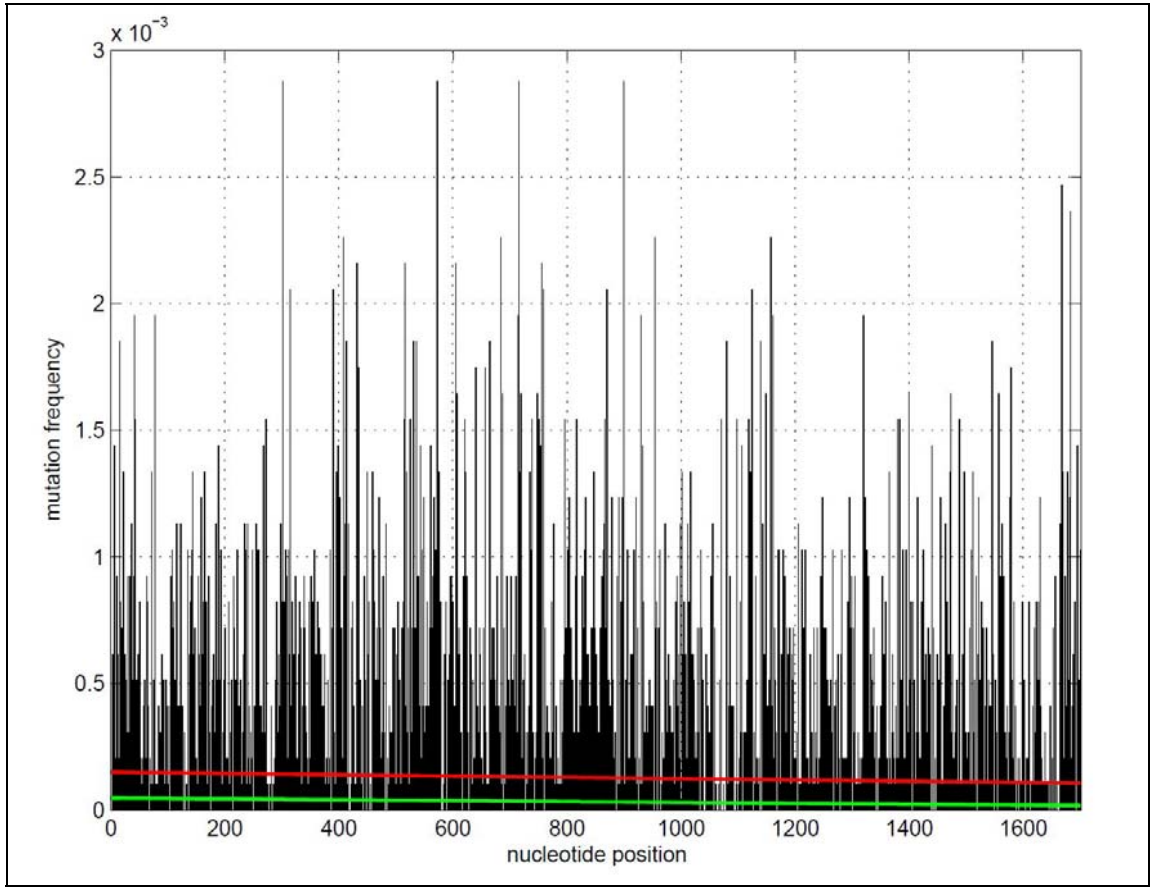

**Figure 5** – The comparison of all mutation frequencies per nucleotide position of 2009 pandemic H1N1 hemagglutinin calculated from the partially disentangled tree (*cf.* **Fig. S2** in supplementary materials) for the whole set of 3243 sequences. The vertical axis corresponds to observed mutation frequencies normalized per sequence per year. The red and green lines show least squares regression fitting for the all mutations (red: $y = -0.768*10^{-6}x + 0.448*10^{-3}$), and the non-equivalent amino acid mutations (green: $y = -0.535*10^{-6}x + 0.142*10^{-3}$), with average mutation rates per sequence per year of $3.83*10^{-4}$ and $9.64*10^{-5}$ respectively.

The dates of each strain collection – which were not used in any way during the tree reconstructions here, and which can be treated only as indicative (over 93% of all 3243 sequences have their full MM-DD-YY isolation timestamp available; see *supra* in the Materials section) – might be used as a further confirmation. They permitted us to verify that indeed the gross majority of descendants had the time stamps succeeding those of their ancestors, validating correctness of the whole approach. In the small 203 set tree (**Fig. 4**) there were just two cases, for which descendant's isolation time stamp preceded that of its ancestor, yielding evolutionary paths with minor chronological failures (A/New_York/3210/2009, April 24 2009 → A/Sao_Paulo/53206/2009 July 19 2009 → A/California/VRDL7/2009 June 17 2009, and A/Scotland/EastKilbride_420874/2009, June 24, 2009 → A/California/VRDL61/2009, July 27 2009 → A/Bogota/WRAIR0442T/2009, June 19 2009).

The **Figure 5** shows the distribution of frequencies for all mutations observed on a partially disentangled tree for the whole 3243 set of sequences. The least squares regressions, along all nucleotide positions, were performed for all mutations as well as for non-equivalent amino acid mutations – in both cases demonstrating almost flat gradients, with a very slight downward slope towards the 3'-end positions. The mean mutation rates per sequence per year were $3.83*10^{-4}$ for the all mutations, and $9.64*10^{-5}$ for the non-equivalent ones. We have also applied the gradient method as used by Fitch (Fitch et al. 1997; their Figure 2); assigning the CY041549 to serve as the assumed root sequence, and counting mutations along branches from such root node to all the terminal leaf sequences. This yielded a slightly smaller average mutation rate per sequence per year of $1.138*10^{-6}$ (with $y = 1.19*10^{-6}x + 4.445*10^{-7}$ as the gradient fit; however, with rather low Pearson correlation of only 0.708).

## *4. Discussion*

We argue that tree reconstructions minimizing a number of spurious multiplication of a single "true" G1171A mutation (to the extent possible by the internal contradictions extant in an analyzed sets of sequences) are important for a better estimation of mutation rates. The **Figure S2** (supplementary materials) shows rectangular phylogram for a the 3243 set – the tree reconstruction required just one SNP to explain that mutation – not surprisingly between the CY041549 and CY107784 sequences, the same as for the 203-sequences tree. Many other, numerously populated, mutations, e.g. the A658T (T221S; with 500 cases of T658), or the A1281G (silent K427; 396 cases) are also predominantly contained within their respective large sub-clades. We have also checked to what extent, a supposedly more stringent, method of maximum likelihood (ML) phylogenetic reconstruction can cope with disentangling the whole 3243 sequences three. The FastTree (Price et al. 2009) program was the obvious choice, as other leading methods for large-scale ML estimation, can require weeks to even months when used on datasets with thousands of molecular sequences. According to Liu et al. (Liu et al. 2011), who performed an extensive testing of existing practical packages, the results from FastTree are of comparable accuracy to other ML methods currently available, at a fraction of the cost in running time. Unfortunately, the resulting solution (**Figure S3**, suppl. materials) does offer no advantage over even the classical NJ-tree reconstruction**:** the resulting sub-branches in both graphs are deeply tangled to a comparable degree.

The actual composition of our large HA set analyzed here determines that there always will be at least one mutation on each of the 1226 positions, irrespective of the tree reconstruction method used. The mutable positions are therefore 72% of the gene length; while 28% are strictly conserved – a proportion in almost exact agreement with Eigen's estimates for RNA viral evolution (Eigen 1993). The ratio of variable *vs.* hyper-variable regions, however, depends on the tree reconstruction approach employed – approximately 40% of these 1226 mutations will have an amino acid changing character, although their exact number and the compositions are dependant to some extent on the reconstruction approach employed. Nevertheless, disentangling tree's branches not only helps to eliminate spurious peaks of overestimated mutation rates, but it also makes possible to trace mutations occurring for the same codons – esp. of non-equivalent amino acid changes subsequent to a previous simple silent nucleotide mutations – both vital factors to consider. Moreover, such accumulation can lead to yet another phenomenon**:** of co-clustering several mutations within the same sub-clades, like e.g. the T1056C (G352, silent mutation, on $3^{rd}$ codon position) with the C1656T (F552, silent, on $3^{rd}$ position), or the T1408C (L470, silent, on $1^{st}$ position) with the C1437T (N471, silent, on $3^{rd}$ position). This might lead to forming a possible "foundation" for a subsequent non-reassortant genetic shift event, of the type recently sought experimentally (Imai et al. 2012, Herfst et al. 2012), and then mathematically modeled by Russell (Russell et al. 2012) in a study of probabilistically estimated substitutions, leading possibly to more virulent strains. But even more important for such model predictions would be the ability to use data for time and location resolved evolution pathways with reasonably better levels of confidence, esp. as our estimated substitution rates for H1N1 pandemic, obtained through the disentangled three method, are of at least magnitude order lower than estimated earlier.

Strelkowa and Lässig (Strelkowa and Lässig 2012) argued, based on 39 years of available serotype H3N2 data, that influenza evolves by a strong clonal interference – a mode of evolution comprising of a race between viral strains with different beneficial mutations, governed not only by positive selection, but also by background selection outside antigenic epitopes, with immune adaptation and conservation of other viral functions interfering with each other. They concluded that a quantitative understanding of influenza's evolutionary and epidemiological dynamics must be based on all genomic domains and functions coupled by clonal interference. The three years worth of data for pandemic H1N1 strains are too short a period to repeat such calculations. However, the vast majority of the more recent isolates are indeed almost exclusively carrying the A1171 form.

Recently an interesting study (Renzette et al. 2012) examined *ab initio* passaging of the A/Brisbane/59/2007 (H1N1), and the A/Brisbane/10/2007 (H3N2) in MCDK cell cultures, followed by a deep sequencing study, and demonstrated that some surprises might await there, as they have shown rather unexpected *increase* in both serotypes' viral diversity – occurring concurrently at the same time, albeit in two separate cell lines. As a deep sequencing *ab initio* experiments produce large amounts of reliable evolutionary data it would be of interest to study them in much more detail. Which pinpoints an obvious advantage of our findings here. The NJ-family of phylogenetic reconstruction approach offers one of the most economic performance levels, and thus augmenting its accuracy levels and reliability might be of a potentially seminal benefit. The preliminary results obtained here by applying our tree disentanglement approach to the whole 3243 collection of unique HA genes are promising, and are currently subject of a more broad in-depth project.

In order to get an independent confirmation of the findings described, after submitting the I-st version of this manuscript, we have examined also the 3243 set by using a globally optimized implementation of the eBURST algorithm – goeBurst (Francisco et al. 2009), as their method does search for a globally optimal solution utilizing a graphic matroid, for which greedy algorithms provide that an optimal solution does exist. The **Figure S3** (supplementary materials) shows the graph for the 3243 set obtained from the goeBurst – the single mutation between the nodes CY041549 and the founder strain CY107784 (aka. the mutation E391K) is prominently visible, providing a strong corroboration indeed.

## *Acknowledgements*

We would like to thank Pat Churchland for looking over the English. This work was partially supported by the EU project SSPE-CT-2006-44405, and also partially supported from the BST/115/30/E-343/S/2012. PP has been partially supported by the framework of European Social Fund through the Warsaw University of Technology development program.

Additional partial funding was generously provided by the WND-POIG.01.01.02-00-007/08 grant from the European Regional Development Fund. 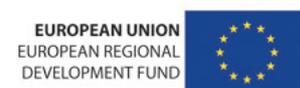

## *Supplementary on-line materials*

## *Appendix*

**The modified NJ+ algorithm**

The distance *D(i,j)* between two sequences *i* and *j* is designated as the number of differences between their nucleotide chains:

$$D(i,j) = \sum_{k=1}^{N} w_{(k)} \left\{ difference \left[ sequence(i,k), sequence(j,k) \right] \right\}, \qquad (1)$$

where N denotes sequence length, and the terms $w_{(k)}$ are nucleotide position specific factors, used to weight a possible influence of mutations at different codon $1^{st}$, $2^{nd}$, and $3^{rd}$ positions. For each position its corresponding $w_{(k)}$ terms might be multiple and different if e.g. at a given position there are mutations of more than one kind (like A→G and A→T, for different sequences). For the $w_{(k)}$ optimization a simple evolutionary algorithm (Chambers et al. 1995) was employed. The initial values of $w_{(k)}$ were determined in dependence whether the nucleotide *k* occupied $1^{st}$, $2^{nd}$ or $3^{rd}$ position in the respective codon as follows:

$$w_{(k)} = A_{(k)}/m_{(k)}, \qquad (2)$$

where $A_{(k)}$ describes how many identical alleles of a given kind are there for the *k* position in the sequences set, and $m_{(k)}$ is a heuristic divisor, different and optimized subsequently by the evolutionary algorithm for each of the *k* $1^{st}$, $2^{nd}$ or $3^{rd}$ position (a good starting point was obtained for the 3243 sequences set with $m_{(k)}$ in the ranges of**:** 8-10 for the $1^{st}$; 12-13 for the $2^{nd}$; and 15-18 for the $3^{rd}$ position, respectively).

The distance between two nodes *(u,v)* in the tree will be denoted as *d(u,v)*. The flow of NJ+ algorithm is similar as in the classic NJ approach, with an addition of few extra steps. At first the number of nodes (r) is equal to the number of sequences, so $d(i,j) = D(i,j)$, the distances between nodes are stored in the matrix $d$. The cost matrix $Q$ is then computed:

$$Q(i,j) = (r-2)d(i,j) - \sum_{k=1}^{r} d(i,k) - \sum_{k=1}^{r} d(j,k), \qquad (3)$$

Then from the $Q$ matrix the entry with minimum value is found, let's note its indexes as $f,g$:

$$(f,g) = \text{argmin}(Q). \qquad (4)$$

In the classical NJ algorithm a new node $u$ is created, and distances $d(f,u)$ and $d(g,u)$ are computed. In our algorithm NJ+ the new node will be created only after we would decide about its creation in the subsequent steps, however the distances $d(f)$ and $d(g)$ are computed same as in the NJ-classic:

$$d(f) = \frac{1}{2}\left( d(f,g) + \frac{\sum_{k=1}^{r} d(f,k) - \sum_{k=1}^{r} d(g,k)}{r-2} \right), \qquad (5)$$

$$d(g) = d(f,g) - d(f).$$

Then the distances $d(f)$ and $d(g)$ are rounded in the way presented below, to assure that edge weight will always represent a full mutations length:

$$d(f), d(g) = \text{DIST\_ROUND}[d(f), d(g)]. \qquad (6)$$

After computing the distances and determining the tree topology, the node character is inferred – that is whether the node will be treated as a leaf or as an internal node. In the NJ-classic the nodes $f$ and $g$ are always assumed to be leaf nodes. The description how we decide on node's character is presented below. In the case of two leaf nodes the algorithm is the same as in the NJ-classic – the new node $u$ is created, and nodes $f$ and $g$ are joined into the $u$ node. The distance between the nodes $u$ and $f$ and $g$ are designated as:

$$d(u,f) = d(f), \qquad (7)$$
$$d(u,g) = d(g).$$

The distance between the new node $u$ and the other (not yet added to the tree) node $k$ is computed as:

$$d(u,k) = \frac{1}{2}\big(d(f,k) + d(g,k) - d(f,g)\big). \qquad (8)$$

Then the nodes $f$ and $g$ are removed from the $d$ matrix, and the new node $u$ is added into the $d$ matrix instead. So far the steps taken were identical to the NJ-classic approach. Note, that the newly created node $u$ has not been assigned sequence's label. In a case that there is one leaf node and one internal node, we need to extend the classic NJ algorithm behavior by adding some new steps. Let's assume that $f$ is an internal node, and $g$ is a leaf. Then, the node $g$ is joined with the node $f$ and it is removed from $d$ matrix. For a case of the $f$ being a leaf and the g an internal node, the steps will be analogous. The last possible configuration of node types to consider will be when both the *f* and the *g* are internal nodes. For such a case we'd need to decide, during their labeling, how we will place them into the tree. If both of them are unlabeled, or if the *f* is labeled and the *g* is unlabeled, then we join with the *f* all the nodes previously joined with the *g*, and then remove *g* from the distance matrix *d*. Notice that *g* will not be present in the final tree, as it was only used as an auxiliary node, to temporarily hold already joined nodes – as in each step only two nodes are being joined. By introducing this step we have dispersed with the limitation for a number of simultaneously joined nodes. In a situation when the *g* is a labeled node, and the *f* is an unlabeled one, we will proceed analogously. In the case when both *f* and *g* are labeled we take the same steps as for the both nodes being terminal leaves. The algorithm steps are summarized in a pseudocode as follows:

```
Repeat until there will be only one node left in the
distance matrix d.
   1. Compute the cost matrix Q.
   2. Find an entry (f,g) with minimum value in the matrix
      Q.
   3. Compute distances d(f) and d(g).
   4. Round distances d(f) and d(g) to full-mutation
      distances.
   5. Determine the nodes' f and g character.
   6. Merge nodes:
       a. If f and g are leaf nodes:
            i. Create a new node u;
           ii. Join nodes f,g with new node u;
          iii. Set d(u,f) = d(f) and d(u,g) = d(g);
           iv. Remove f and g from the distance matrix d;
            v. Add u to the distance matrix d, and update
               the matrix d values.
       b. If f is an internal node and g is a leaf:
            i. Join node g with node f;
           ii. Remove g from the d matrix.
       c. If g is an internal and f is a leaf:
            i. Join node f with node g;
           ii. Remove f from d matrix.
       d. If f and g are internal nodes:
            i. If f is labeled and g is unlabeled or if
               both are unlabeled:
                1. Move to f all previously joined nodes
                   of g;
                2. Remove g from d matrix.
           ii. If g is labeled and f is unlabeled:
                1. Move to g all previously joined nodes
                   of f;
                2. Remove f from d matrix.
          iii. If both are labeled:
                1. follow the steps same way as for the
                   6a.
```

**Distance rounding**

The distance between two sequences is an integer representing a number of differences between their nucleotides chains. However, the distances computed in the equation (3) are real numbers. To assure that all edges in the tree represent a number of mutations between sequences we round their distances. The DIST_ROUND function is described in the pseudocode as follows:

```
d(f), d(g) = DIST_ROUND(d(f), d(g)):
Sum = d(f) + d(g);
// {d(f)} is a fractional part of d(f)
// round(d(f)) is the usual round function
if( {d(f)} > {d(g)} ) then
     d(g)=round(d(g))
     d(f)=Sum-round(d(g))
else
     d(f)=round(d(f))
     d(g)=Sum-round(d(f))
end;
```

Let's see how it works on an example, for *d(g)=3.2* and *d(f)=3.4* the DIST_ROUND function will return *d(g)=3* and *d(f)=4*.

**Determination of node's character**

Determination of node's character is decided by evaluating two conditions during internal nodes searching. First, the node's edge length *d(f)* and *d(g)* is examined, if edge length is smaller than one (which corresponds to exactly one mutation), then the node is treated as an internal node, otherwise it is a leaf. If, by fulfilling this condition, nodes are assigned to be leaves, the second condition is used to ensure that any of such node is not an internal node. Second condition checks triangular equality between three node sequences. Let's assume that the *f* and *g* are labeled nodes. When *f* or *g* have not a label, the label for them is taken from theirs joined nodes – as the label of the closest labeled node. When, as in creating a third labeled node, we are about to select the node *h*, which is either the closest to the *f* or to the *g*, and which is still present in the distance matrix *d* – then for the candidate selected nodes we have their corresponding sequences. Lets denote them as F, G, H. We need then to check whether their distances will satisfy the equation:

$$D(F,H)=D(F,G)+D(G,H). \qquad (9)$$

If the equation (9) is satisfied then node G is an internal node, and the node F will be joined with G. If equation (9) is not satisfied the analogous condition is checked, which assume the F as being a possible internal node.

*Supplementary Figures and Tables*

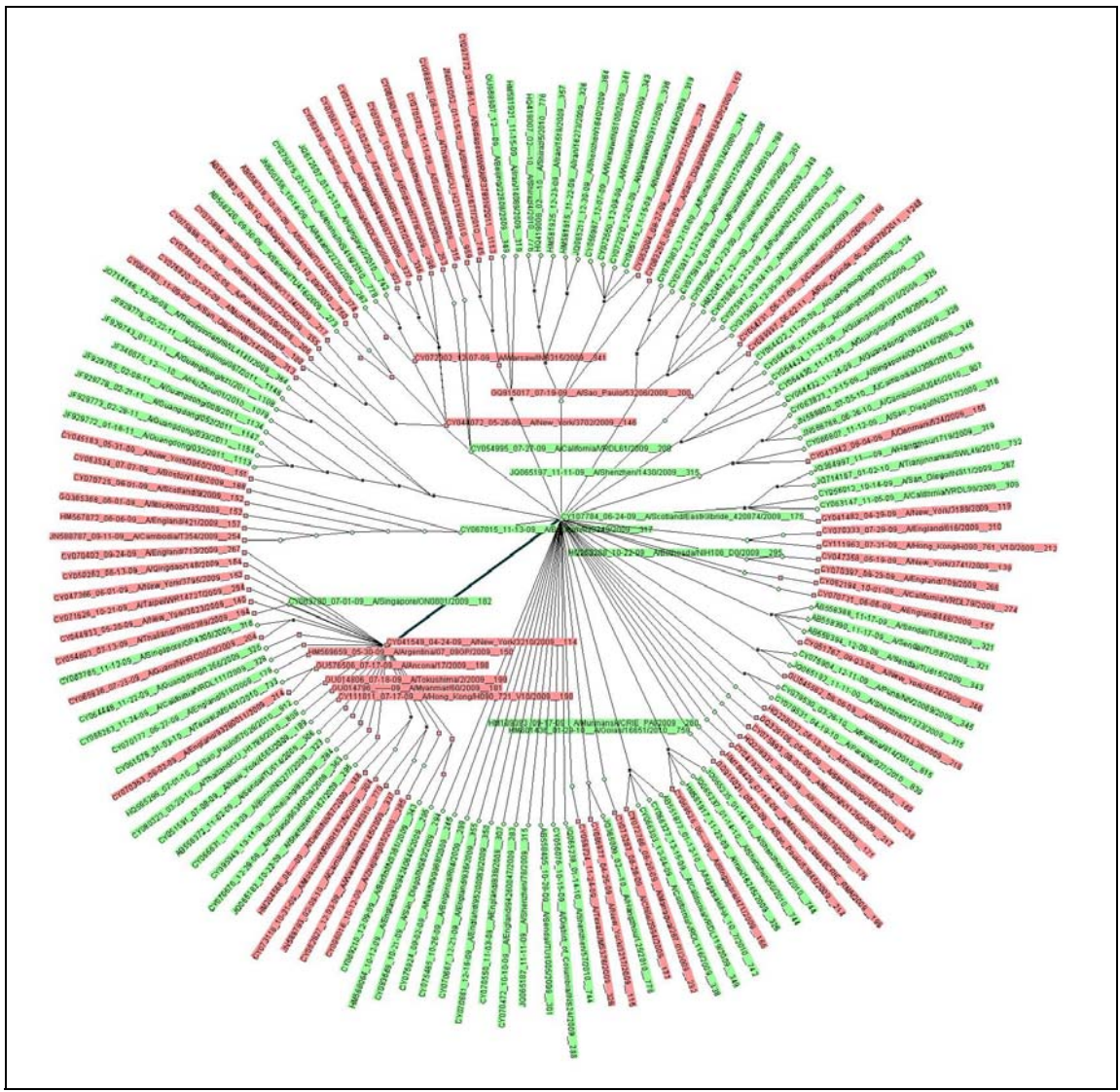

**Figure S1** – The NJ+ circular phylogram for the 2009 pandemic virus H1N1 hemagglutinin 203 sequences set (*see the main text for explanation of this set and the tree origin*). Each of sequence labels comprises of four parts, e.g. for the CY041549_04-24-09__A/New_York/3210/2009__114 there is the accession #, isolation time stamp in the MM-DD-YY format, the strain name, and the number of days that passed since the Jan 1$^{st}$ 2009 till the day of sample isolation. Some labels of internal nodes are not displayed to preserve clarity. A zoomable Fig_S1.pdf file in Supplementary materials should be used for a better visibility of details.

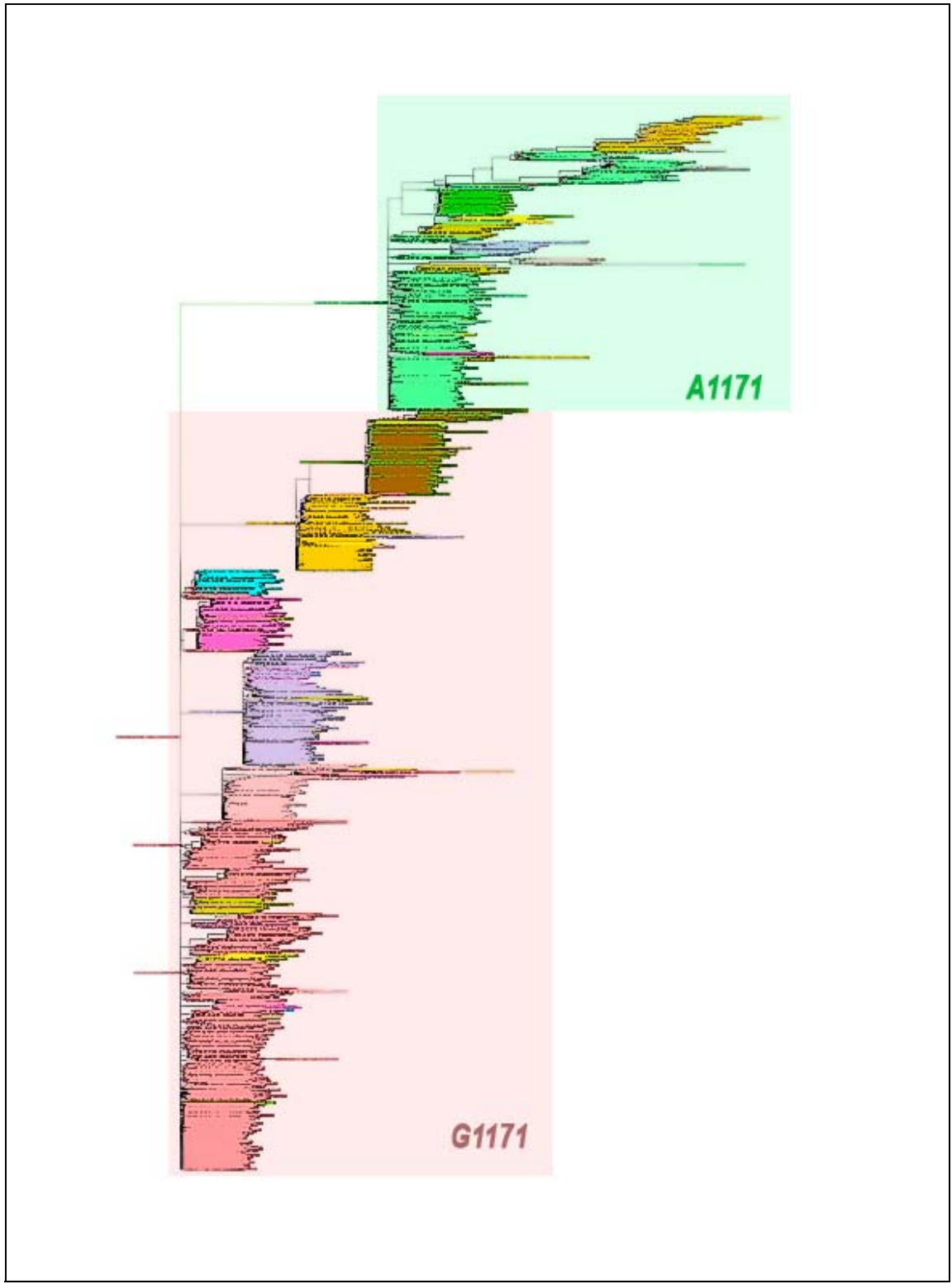

**Figure S2** – The partially disentangled NJ+ phylogram of the 3243 set of 2009 pandemic H1N1 hemagglutinin 1701 NN sequence. The zoomable Fig_S2.nexml file can be used for a better visibility of details. The NEXML phylogenetic graph files can be viewed using the Dendroscope software of Huson et al. (BMC Bioinformatics 8:460, doi:10.1186/1471-2105-8-460). The color schemes used to distinguish sub-clades and sequences are given in the **Table ST1)**

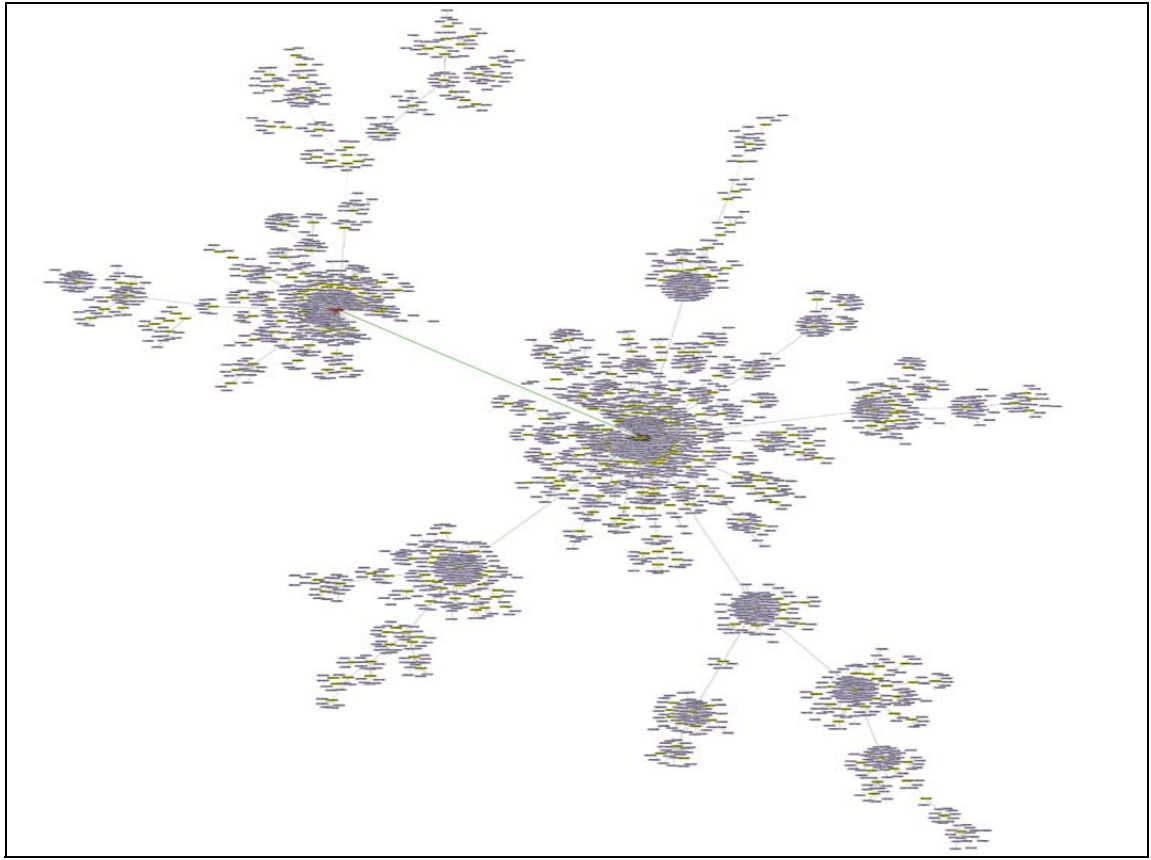

**Figure S3** – The visual representation of the results from the globally optimized eBURST approach (Francisco et al. 2009) for the 3243 set of 2009 pandemic H1N1 hemagglutinin, full length 1701 NN sequences. The zoomable Fig_S3.pdf file can be used for a better visibility of details. The E391K mutation, between sequences CY041549 and the founder strain CY107784 is marked by a thicker, green line (from approx. the middle, towards top-left)**.** Unfortunately, the current implementation of goeBurst doesn't provide an option to output the results in the Newick format, and therefore an automatic *in silico* calculation of site-specific mutation rates from the goeBurst is not yet possible (as part of the subsequent project we are currently working on providing additional functionalities to the goeBurst results).

| nucleotide mutation | position in codon | amino acid mutation | mutation cluster labels |
|---|---|---|---|
| G1171A | 1st | **E391K** | G1171 (squares); A1171 (circles) |
| A658T | 1st | T221S | HQ695920_06----09__A/Taiwan/14/2009 |
| A1281G | 3rd | K427 (silent) | JN393307_10-21-09__A/Hue/15/2009 |
| T1408C | 1st | L470 (silent) | GQ219579_05-16-09__A/Osaka/2/2009 |
| T1056C | 3rd | G352 (silent) | JQ396230_07-08-11__A/KENYA/131/2011 |
| G1403A | 2nd | S468N | *partially co-clustering with the* T1056C |
| G605C | 2nd | S202T | *partially co-clustering with the* T1056C |
| G340A | 1st | D114E | *partially co-clustering with the* G1403A |
| T717A | 3rd | D239E | CY070602_11-11-09__A/Scotland/79/2009 |
| G640A | 1st | A214T | CY045226_06----09__A/Taiwan/115/2009 |
| T433C | 1st | S145P | CY095878_11----09__A/Hubei/76/2009 |
| C145A | 1st | L49I | CY062931_12-18-09__A/Berlin/INS171/2009 |
| G1012A | 1st | V338I | CY050273_09-03-09__A/Qingdao/477/2009 |
| C1437T | 3rd | N471 (silent) | *partially co-clustering with the* T1408C |
| C144T | 3rd | N48 (silent) | CY062867_12-18-09__A/Athens/INS160/2009 |
| G930T | 3rd | Q310H | CY072950_09-29-09__A/Managua/5258.02/2009 |
| C1656T | 3rd | F552 (silent) | *partially co-clustering with the* T1056C |
| A4G | 1st | K2Q | GQ287625_06-02-09__A/Tokushima/1/2009 |
| A478G | 1st | G160S | *partially co-clustering with the* G605C |
| T852C | 3rd | I284 (silent) | CY103934_09-24-09__A/Pernambuco/624/2009 |
| A716G | 2nd | D239E | HM581925_12-23-09__A/Iran/1519/2009 |
| G665A | 2nd | R222K | GQ280264_06-18-09__A/Nanjing/1/2009 |
| A1159C | 1st | N387H | GU576540_12-31-09__A/Ancona/508/2009 |
| T1668C | 3rd | S556 (silent) | HM189390_11-30-09__A/Perm/CRIE_ZTS/2009 |
| A424G | 1st | N142D | CY060582_10-24-09__A/Ontario/304434/2009 |

**Table ST1** – the 25 most populated mutation positions of the 3243 set of 2009 pandemic H1N1 hemagglutinin sequences; the last column shows the labels on the phylogram available as a zoomable SVG file for the **Figure S5.** All strains of G1171 have their nodes marked as squares, and all of A1171 ones with circles of varying size. Other mutations are marked by different combinations of their nodes, label colors and font styles, as wall as differently sized node markers.